# THE NON-LINEAR RESPONSE OF THE MAGNETOSPHERE: 30 OCTOBER 1978

C. P. Price and D. Prichard

Physics Department, University of Alaska Fairbanks

*Abstract.* Previous efforts to find evidence of deterministic nonlinear dynamics in the global geomagnetic system have treated the geomagnetic system as autonomous. However, the geomagnetic system is strongly driven by the stochastic solar wind. We consider the response of the magnetosphere, as given by the AE index, for one day when the IMF had a nearly constant southward value. Using both a series of non-linear statistics and non-linear prediction of the response to the input signal $vB_s$, we find that there is some evidence for deterministic non-linear response of the Earth's magnetosphere on that day.

## Introduction

There have been numerous studies [*e.g.* Clauer *et al.*, 1981; Baker *et al.*, 1983; Bargatze *et al.*, 1985; Tsurutani *et al.*, 1985, 1990; Gonzalez *et al.*, 1989; Vassiliadis *et al.*, 1992b] trying to relate solar wind parameters to geophysical indices (*e.g.* AE, AL, $K_p$, *etc.*) These studies show that the magnetospheric system is neither periodic nor quasiperiodic, and that the magnetosphere acts as a type of low-pass filter on the solar wind input. This suggests the possibility of a non-linear magnetospheric response. These studies also show that only *part* of the AE variation can be explained as being a direct linear response to the solar wind.

Recently there has been much interest in determining whether the Earth's magnetospheric system behaves like a low dimensional (*i.e.* deterministic) non-linear system [Vassiliadis *et al.*, 1990; Shan *et al.*, 1991a,b; Roberts *et al.*, 1991; Roberts, 1991; Vassiliadis *et al.*, 1991; Prichard and Price, 1992]. These studies, which focus on calculating the properties of a possible strange attractor in the Earth's magnetosphere, do not yield clear evidence for deterministic non-linear behavior, although some non-linear circuit analogue models show promising results [Baker *et al.*, 1990; Klimas *et al.*, 1992].

All of the previous non-linear studies of the magnetospheric response have focused solely on the state of the magnetospheric system. There are problems with this approach. First and foremost, calculations intended to discover the properties of a strange attractor (*viz.* the fractal dimension) assume tacitly that the system is either autonomous or periodically driven. When a system (such as the Earth's magnetosphere) is driven stochastically, the trajectory in phase space is usually not able to converge to an attractor. Thus, attempts to determine the non-linear properties of the system solely from the response of the system (no matter how complete the diagnostics are) will not be meaningful, since the strange attractor is in general never delineated. Clearly, it is imperative to consider simultaneously the solar wind driver, and the only way to do so in general is to use non-linear input/output prediction techniques [Farmer and Sidorowich, 1988; Casdagli, 1989; Eubank and Farmer, 1990; Casdagli, 1992a,b; Hunter and Theiler, 1992]. Second, since the results of non-linear calculations of any kind on real data sets frequently give less than clear-cut results, it behooves one to have a standard of comparison, for which we will use surrogate data sets [Theiler *et al.*, 1992].

In this paper, we analyse the response of the magnetosphere, as given by the AE index, for the day 30 October 1978, when the interplanetary magnetic field (IMF) had a nearly constant southward value. Our choice is motivated by the following reason: only for periods when the forcing function is relatively steady, *i.e.* the input function reduces to a point in state space, is it possible for the trajectories in phase space to have time to converge onto a strange attractor. Thus, this interval may be seen as a bridge between the previous analyses and non-linear prediction analyses. We present the results of two distinct analyses of the magnetospheric response: first, we make the standard dimensional estimates (via correlation integral and local singular value decomposition), and second, we make a non-linear prediction of the response to the input signal $vB_s$.

## Data Sets and Methods of Analysis

The response of the magnetosphere is taken to be given by the auroral electrojet index AE, with one minute resolution. Although there are problems with this choice [see *e.g.* Kamide and Akasofu, 1983; Prichard, 1993], the use of the electrojet indices has been standard for this type of investigation. Here, the solar wind input is taken to be the product of the southward component of the IMF ($B_s = \frac{1}{2}[|B_z| - B_z]$) and the solar wind speed, calculated from ISEE-3 measurements [Ogilvie *et al.*, 1978] with one minute resolution. Figure 1 shows the solar wind input $vB_s$ and the magnetospheric response AE time series for 30 October 1978; note during a large part of the day the IMF had a nearly constant value of $B_z = -10$ nT. For the non-linear prediction, we will use the data of the entire month October 1978. Although the 30 October 1978 AE data set length is rather short (1440 elements) for dimension estimation [Eckmann and Ruelle, 1992] we are using the dimension estimates as convenient non-linear statistics (see below).

We have also constructed surrogate data sets (nondeterministic sets with the same statistical properties, *viz.* power spectrum, autocorrelation, *etc.*) for the AE data in order to provide a standard for comparison. Surrogates can easily be made by taking the Fourier transform, randomizing the phases, and then inverting the transform. Our surrogates also have the same amplitude distribution in the time domain as the original data [Theiler *et al.* 1992]. We employ the surrogate data sets to determine the significance of the results of our analyses. If the results of the non-linear statistics (here, dimension estimates) for the AE data set vary meaningfully from the results for the surrogate data sets, one may assume the presence of a deterministic non-linear process in the magnetosphere.

We assume that the signal (*e.g.* AE) is due to a non-linear dynamical process involving an unknown number of variables. Using the method of time delays (Packard *et al.*, 1980; Takens, 1981) we reconstruct the attractor by creating a set of $D_E$ dimensional vectors from the original time series $\{x(i)\}$: $\vec{x}_i = [x(i), x(i+\tau), x(i+2\tau), \dots, x(i+(D_E-1)\tau)]$ with $\tau$ the time delay, and $D_E$ the embedding dimension. This reconstruction will have the same topological and dynamical properties as the original attractor as long as $D_E > 2D$, where $D$ is the attractor dimension (Sauer *et al.*, 1991.)

One estimate of the attractor dimension uses the scaling of the number of points inside a ball as the radius of the ball (r) increases. One computes the correlation integral [Grassberger and Procaccia, 1983]:

$$C(r) = \frac{2}{(N-W)(N-W+1)} \sum_{j=W}^{N} \sum_{i=1}^{N-j} \Theta\left(r - \|\vec{x}_{i+j} - \vec{x}_i\|\right)$$

Removing W points on either side of the reference point eliminates autocorrelation effects [Theiler, 1986]; for $W = 1$ we recover the standard correlation integral. $W$ is on the order of

30 October 1978

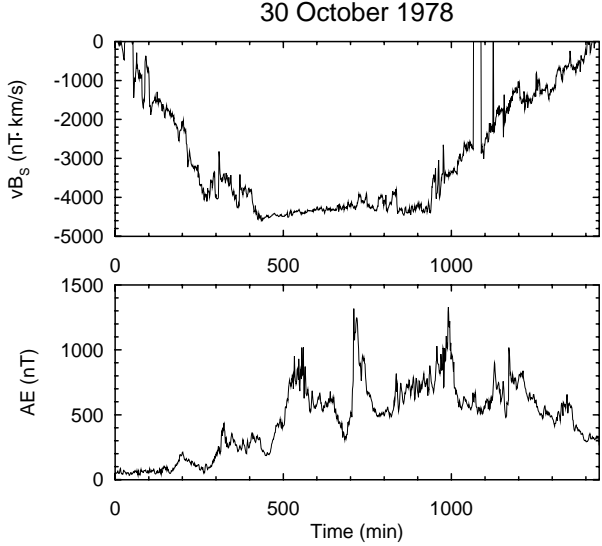

Fig. 1. The time series for the solar wind input $vB_s$ and magnetospheric response AE for 30 October 1978.

the autocorrelation time. $C$ should scale (over a limited range) as: $C(r) \sim r^d$ where $d$ is the dimension. Takens [1981] suggested a direct dimensional estimate based on the $M$ distances less than some upper cutoff $r_0$,

$$D_{Takens} = -\left[\frac{1}{M}\sum_{i=1}^{M} log(r_i/r_0)\right]^{-1}$$

We choose $r_0$ to be roughly $\frac{1}{2}$ the rms deviation of the time series; the Takens estimate can be calculated directly from the correlation integral [Theiler, 1990]. Another non-linear statistic closely related to the correlation integral is the BDS statistic: $BDS = \sqrt{N}[C(r;m) - C(r;1)^m]$ where $C(r;m)$ is the correlation integral evaluated at radius $r$ in $m$ embedding dimensions [Brock, 1988]. For stochastic time series, $C(r;m) = [C(r;1)]^m$, so the deviation of the $BDS$ statistic from zero gives a measure of the determinacy of the time series.

Another dimension estimator is local singular value decomposition [Broomhead et al., 1987; Passamante et al., 1989]. The nearest $n$ neighbors to a reference point form a hyperellipsoid with $m$ significant principal axes; $m$ is the local dimension of the attractor manifold. Averaging $m$ over the reference points gives the dimension estimate $\tilde{D}_{svd}$. The choice of delay is almost arbitrary. The $n \times d_E$ trajectory matrix

$$X = \begin{pmatrix} \vec{\delta x}_1 \\ \vec{\delta x}_2 \\ \vdots \\ \vec{\delta x}_n \end{pmatrix} = \begin{pmatrix} \delta x_1 & \delta x_{1+\tau} & \cdots & \delta x_{1+(D_E-1)\tau} \\ \delta x_2 & \delta x_{2+\tau} & \cdots & \delta x_{2+(D_E-1)\tau} \\ \vdots & \vdots & \ddots & \vdots \\ \delta x_n & \delta x_{n+\tau} & \cdots & \delta x_{n+(D_E-1)\tau} \end{pmatrix}$$

is used to form the covariance matrix $Z = X^T X$. Here $\vec{\delta x}_j$ is the difference between the point $\vec{x}_j$ and the reference point. $m$ is estimated by the number of significant eigenvalues of $Z$.

The best approach is to consider the magnetosphere as a non-linear input/output system, that is, to simultaneously examine both solar wind parameters and geomagnetic indices, and compare the output to a non-linear predictor [Farmer and Sidorowich, 1988; Casdagli, 1989; Eubank and Farmer, 1990; Casdagli, 1992a,b; Hunter and Theiler, 1992]. For a driven system with input $u(t)$ and output $y(t)$ we make embedding vectors $\vec{c}(t) = [u(t), u(t-\tau)\ldots u(t-(D_{E1}-1)\tau), y(t), y(t-\tau)\ldots y(t-(D_{E2}-1)\tau)]$. We use the first 29 days of October 1978 to "train" the phase space $\{\vec{c}\}$. To predict $y(t+1)$ we first find the $k$ nearest neighbors to $\vec{c}(t)$ in the state space, then we determine where these points go after 1 time step, and then find a least squares solution for $y(t+1)$.

To determine the quality of the prediction, we randomly choose 50 points in the state space during 30 October 1978, and predict the value 1 step ahead. After all 50 predictions are made we compute the normalized error $E_T$:

$$E_T^2 = \frac{1}{50}\sum_{i=1}^{50}\left(\frac{y_{predicted} - y_{real}}{\sigma}\right)^2$$

with $\sigma$ the standard deviation of the time series. As $k$ approaches the "training" set length this method becomes a linear predictor. If there is a non-linear relationship between input and output then there will be a minimum of $E_T$ for $k$ small compared to the length of the "training" set [Casdagli, 1992a].

### Results

We have examined non-linear statistics for the 30 October 1978 AE data set and for ten surrogates to that set. Figure 2 shows the dependence on the Takens dimension estimate $D_{Takens}$ as a function of the embedding dimension $D_E$ for the AE data set (solid line) and for the surrogates (points). The delay time $\tau$ is 15 minutes, and $W$ is 50 minutes. We caution the reader from drawing the inference from this figure that the AE data set shows a dimension of roughly 3; in fact, the slopes of the correlation integrals are neither convergent as $D_E$ is varied nor flat at any radius. We remind the reader that we are using the Takens dimension estimate as a non-linear statistic, not as a measure of the effective degrees of freedom. What Figure 2 does show is a statistically significant difference between the AE data set and the ensemble of surrogates.

Figure 3 shows the BDS statistic as a function of radius for the AE data set (solid line) and for the surrogates (dashed lines). The delay time $\tau$ is 15 minutes, $W$ is 50 minutes, and the embedding dimension $m = 10$. Figure 3 also shows a statistically significant difference between the AE data set and the ensemble of surrogates.

N=1440 min., τ=15 min., W=50 min., $r_0$=200 nT

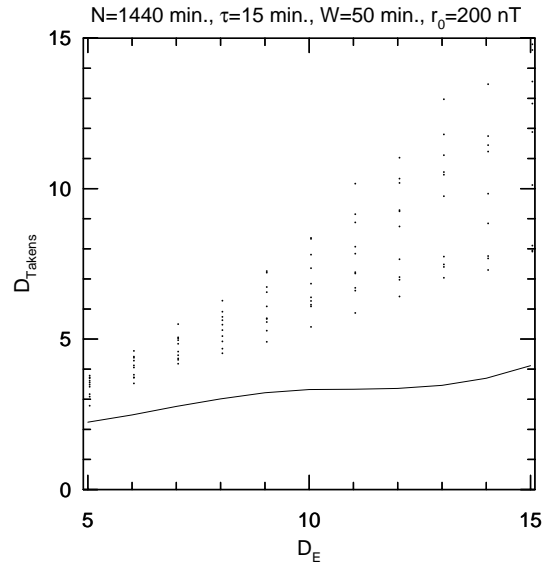

Fig. 2. The Takens dimension estimate $D_{Takens}$ versus the embedding dimension $D_E$ for the AE data set (solid line) and for the surrogates (points). The delay time is $\tau = 15$ minutes, and the value of $W$ is 50 minutes.

N=1440 min., τ=15 min., W=50 min., D_E=10

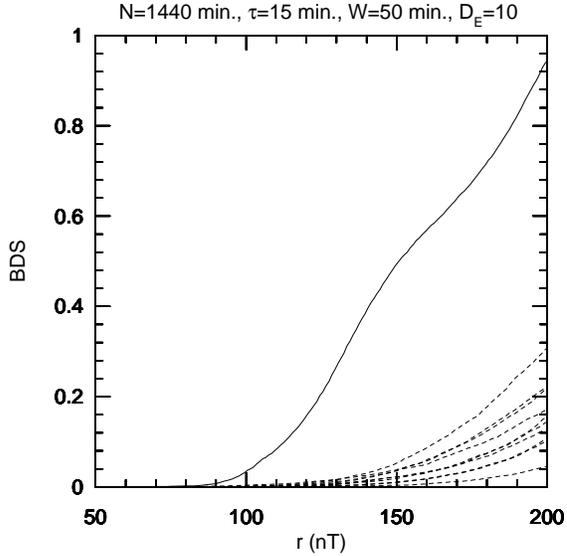

Fig. 3. The BDS statistic as a function of radius for the AE data set (solid line) and for the surrogates (dashed lines). The delay time is $\tau = 15$ minutes, the value of $W$ is 50 minutes, and the embedding dimension $m = 10$.

Figure 4 shows the singular value decomposition dimension estimate $D_{svd}$ as a function of the number of neighbors for the AE data set (solid line). The delay time $\tau$ is 1 minute, the embedding dimension is $D_E = 15$ and the least significant eigenvalue is chosen arbitrarily to be 50 times smaller than the largest eigenvalue. The results are not affected significantly by variations in the embedding dimension (we have calculated out to $D_E = 60$) or the least eigenvalue size. We caution the reader from inferring from this figure that the AE data set shows a dimension of roughly 7 (or that the stochastic surrogates show a dimension of roughly 10!). Again, we are using the singular value dimension esti-

N=1440 min., τ=1 min., W=50 min., D_E=15

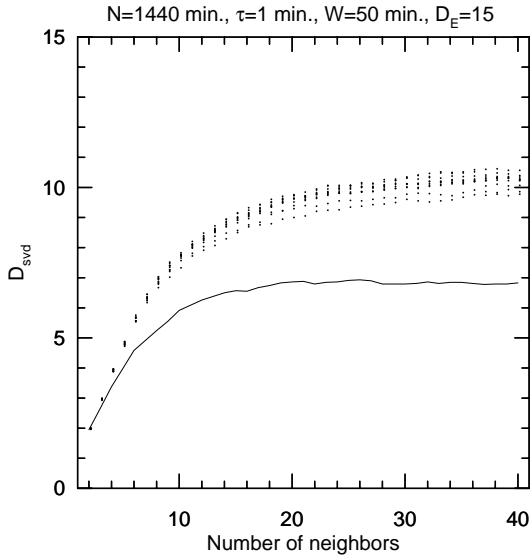

Fig. 4. The singular value decomposition dimension estimate $D_{svd}$ versus the number of neighbors for the AE data set (solid line) and for the surrogates (points). The delay time is 1 minute, the embedding dimension is $D_E = 15$, the value of $W$ is 50 minutes, and the least significant eigenvalue is 50 times smaller than the largest eigenvalue.

N_train=30000, τ=15, D_E1=D_E2=8

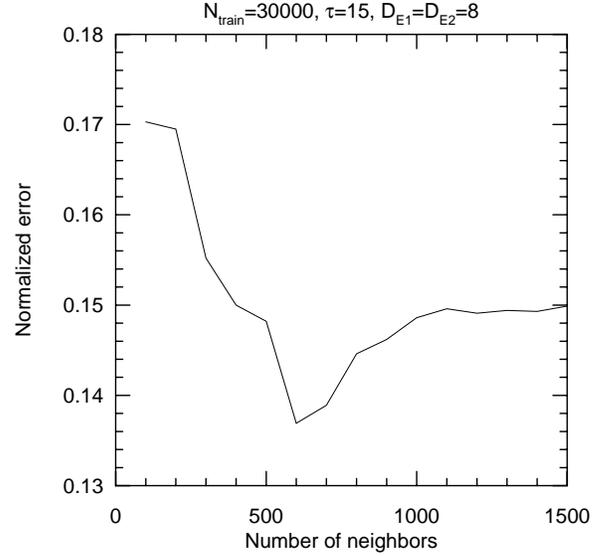

Fig. 5. Normalized one step ahead prediction error ($E_T$) versus the number of neighbors. The delay time is $\tau = 15$ minutes, and the embedding dimension is $D_E = 8$ for both the input and output data sets. The training interval for the nonlinear predictor contains 30000 data points.

mate as a non-linear statistic, not as a measure of the effective degrees of freedom. What Figure 4 does show is a statistically significant difference between the AE data set and the ensemble of surrogates.

In Figure 5 we show the normalized prediction error versus the number of neighbors used to compute the local maps. The delay time $\tau$ is 15 minutes and the embedding dimensions are $D_{E1} = D_{E2} = 8$. The normalized error shows a distinct minimum of $E_T = 0.137$ at roughly 600 neighbors (out of a 30000 point training set). For very large numbers of neighbors (i.e. a linear predictor) the normalized error approaches $E_T = 0.15$. Therefore, the non-linear predictions are roughly 10% better than linear predictions.

## Conclusions

On the day 30 October 1978, the IMF had a nearly constant southward value of $B_z = -10$ nT, and the magnetospheric activity (as indicated by the AE index) was quite high. Although the magnetosphere is not an autonomous system, when the input forcing function is relatively steady, the system may have time to converge to an attractor. Hence, while in general the magnetosphere must be analysed as a non-linear input/output system, for this day (when the input function is approximately a point in state space) we can use both the standard dimensional estimates and non-linear prediction methods.

Our approach has been solely to determine whether or not the magnetosphere gave evidence of a deterministic non-linear response on that day. We are not yet attempting to quantify the non-linearity. Using convenient non-linear statistics, we have compared the AE data set to an ensemble of surrogate data sets. In the same vein, the non-linear prediction analysis is a comparison of the non-linear prediction to a linear prediction of the output AE signal from the input signal $vB_s$.

We have found that the non-linear statistics show significant differences between the AE data set and the ensemble of surrogates. Further, we find that the non-linear predictor is somewhat better than a linear predictor. The combination of these results leads us to conclude that, for 30 October 1978, there is evidence for deterministic non-linear coupling between the solar wind and the magnetosphere. Further work is needed to identify the nature and characteristics of that coupling.

Acknowledgements. We wish to thank J. Theiler, N. Hunter, D. Vassiliadis, A. Klimas, S. Sharma and T. Chang for many useful discussions. We thank NSSDC for assistance in acquiring the ISEE-3 data, and acknowledge S. J. Bame (plasma) and E. J. Smith (magnetometer) as Principal Investigators on those ISEE-3 instruments. One of us (D. P.) would like to acknowledge the support of a UAF Chancellors Graduate Fellowship and an AWU/DOE Thesis Parts Fellowship. This work is supported by NSF grant ATM-9213522.

C. P. Price and D. Prichard, Physics Department, University of Alaska Fairbanks, AK 99775